RESEARCH PAPER

# The Edge Sensor of Segmented Mirror Based on Fringes of Equal Thickness

View the article online for updates and enhancements.

# The Edge Sensor of Segmented Mirror Based on Fringes of Equal Thickness

Xi Zhang[1,2,3], Heng Zuo[1,2], Yong Zhang[1,2,4,5], and Ye-Ping Li[1,2]
[1] Nanjing Institute of Astronomical Optics and Technology, National Astronomical Observatories, Chinese Academy of Sciences, Nanjing 210042, China; hengz@niaot.ac.cn
[2] Key Laboratory of Astronomical Optics and Technology, Nanjing Institute of Astronomical Optics and Technology, Chinese Academy of Sciences, Nanjing 210042, China
[3] University of Chinese Academy of Sciences, Beijing 100049, China
[4] National Astronomical Observatories, Chinese Academy of Sciences, Beijing 100101, China
[5] Key Laboratory of Optical Astronomy, National Astronomical Observatories, Chinese Academy of Sciences, Beijing 100101, China

## Abstract

Co-phase and co-focus detection is one of the key technologies for large-aperture segmented mirror telescopes. In this paper, a new edge sensor based on fringes of equal thickness is developed, which can detect each segment's relative piston, tilt, and tip errors from the interferograms. Based on the co-focus demand for many ground-based seeing limited segmented mirror telescopes, an edge sensor prototype based on such a principle is built and applied in the indoor segmented mirror experiment system in the lab. According to the co-focus requirement of the Large Sky Area Multi-Object Fiber Spectroscopic Telescope, many simulations and experiments are carried out for co-focus error detection of the segmented mirror system. Experiment results show that the co-focus accuracy is better than $0''\!.02$ rms, which can meet the co-focus requirements of most large or extremely large segmented mirror astronomical telescopes.

*Key words:* instrumentation: interferometers – astronomical instrumentation – methods and techniques – techniques: interferometric

## 1. Introduction

Active optics technology is crucial for large aperture segmented mirror telescopes to achieve maximum resolution and the strongest light-gathering power. Co-phase calibration and maintenance are frequently performed before and during the astronomical observation. For co-phase calibration, it can be divided into two steps: co-focus and co-phase. During the co-focusing step, the tip/tilt error between the mirror segment is measured and adjusted using Shack-Hartmann wave front sensors. Furthermore, during the co-phase step, the piston error is measured and adjusted to reach the perfect diffraction-limited performance of the full aperture optical system (Jacob & Parihar 2015). Various detection methods have been studied to detect the relative displacement(piston) and attitude(tip/tilt) between the neighboring segments, such as the Broadband/Narrowband Phasing algorithms (Chanan et al. 1998, 2000), Pyramid Phase Sensing method (PYPS) (Pinna et al. 2008), Dispersed Fringe Sensing method (DFS) (Shi et al. 2004), and Phase Retrieval (PR) (Acton et al. 2012). The broadband/narrowband phasing algorithms have been successfully used for the Keck telescope, the dispersed fringe sensing method has been used in the coarse phasing of the James Webb Space Telescope (JWST), and the phase retrieval used in the fine phasing of JWST. During the co-phase calibration stage, the optical detection methods above are usually adopted to detect the segment errors and set the zero-point for edge sensors; in the co-phase maintenance stage, edge sensors are used to detect the piston error of segments to maintain the co-phase state.

The Large Sky Area Multi-Object Fiber Spectroscopic Telescope (LAMOST, Cui et al. 2012), a special quasi-meridian reflecting Schmidt telescope in operation for 12 yr. The optical system of LAMOST consists of a reflecting Schmidt Ma, a spherical primary mirror Mb and a focal surface in between. Mb has a size of 6.67 m × 6.05 m, which consists of 37 hexagonal spherical mirrors, each of them having a diagonal diameter of 1.1 m and a thickness of 75 mm. During the observation, the spherical mirror Mb is fixed, so there is no obvious change in gravitational force exerted on its mirror support and co-focus system, as is shown in Figure 1 (Cui et al. 2012). Currently, LAMOST has realized the co-focus of the segmented mirror telescope in the visible band based on active optics technology, including calibration with a wave front sensor and adjustment with actuators. Edge sensors are required for Mb to maintain the co-focus during the operation, and this edge sensor prototype is designed for the co-focus of Mb. Many ground-based large segmented mirror telescopes similar to LAMOST are working in seeing limited observation mode without adaptive optics, such as the Hobby–Eberly Telescope (HET) (Palunas et al. 2004) and the Southern African Large Telescope (SALT) (Buous et al. 2008). It is essential for these telescopes to adopt only co-focus active optics to achieve high light-gathering power and efficiency. Edge sensors play a crucial role in providing the active optics system with real-time fine co-focus error detection of the segmented mirror.

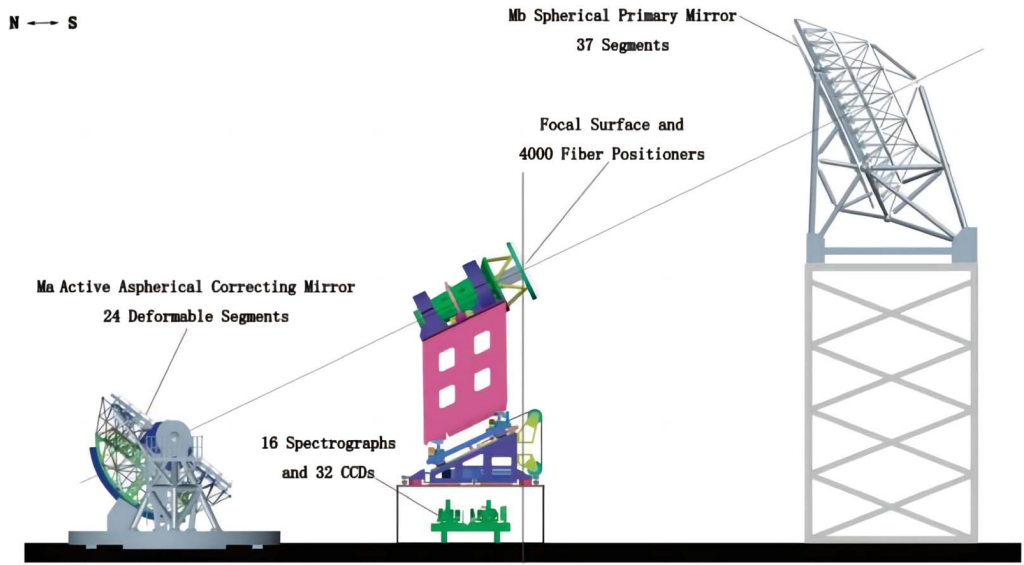

**Figure 1.** Configuration of LAMOST. Image reproduced from Cui et al. (Cui et al. 2012).

Currently, the edge sensors for segmented mirrors mainly adopt electrical measurement methods, including capacitive, inductive, or eddy current sensors. These sensors are applied on various large segmented mirror telescopes in the world. Thirty Meter Telescope (TMT) (Mast et al. 2006), Keck telescope (Jared et al. 1990), and Gran Telescopio Canarias (GTC) telescope use similar capacitive edge sensors (Lefort & Castro 2008); Extremely Large Telescope (ELT) will uses inductive edge sensors (Wasmeier et al. 2014). The parameters of each edge sensor are as follows: the capacitive sensor of the Keck telescope has a measurement accuracy of 3 nm, a range of $\pm 12\ \mu m$, with a temperature drift of 2 nm $°C^{-1}$ and a time drift of 3.2 nm per week (Minor et al. 1990); the TMT telescope's edge sensor was improved from the Keck one with a measurement accuracy of 1 nm, but still has a temperature drift (Mast et al. 2006); the inductive edge sensor for ELT has a measurement accuracy of 1 nm, a range of 400 $\mu m$, and a temperature drift of 1.32 nm $°C^{-1}$ (Rozière et al. 2008). The measurement accuracy of such electrical sensors is greatly affected by the environment, such as temperature drift and time drift, which are challenging to eliminate, and have poor anti-electromagnetic interference ability. Due to the temperature drift phenomenon, the electrical edge sensors mentioned above must be calibrated periodically. For LAMOST, the impact of such environmental temperature changes will be more obvious. This is because the temperature range at most Chinese astronomical observatories is above $50°C$ per year, at least twice that of other sites around the world. It is time to find another candidate edge sensor different from traditional electrical edge sensors.

Based on the edge sensor detection principle of the interference fringes of equal thickness proposed by our team earlier (Zuo et al. 2021), combined with the indoor segmented-mirror active optics experiment system (Su et al. 2000), an optical edge sensor experimental prototype for co-focus detection of tip/tilt errors is developed and a new interferogram detection algorithm is proposed. The advantages of this optical edge sensor are: the segment errors are measured directly, which simplifies the process of co-phase and co-focus without the need for calibration; the sensing method based on fringes of equal thickness is not susceptible to temperature changes as electrical edge sensors, therefore the temperature drift is eliminated completely. With this sensor, the tilt/tip/piston error of the segments can be obtained and used as high-precision feedback to provide real-time correction for actuators, which enables co-phase maintenance. Simulation and experiments were performed on the integrated system, and the co-focusing performance was achieved within the technical requirements.

In the following section, the detection principle and the optical system structure of the edge sensor are described. The interferogram processing algorithms based on fringes of equal thickness are introduced in Section 3. The simulation and experiment tests are described separately in Sections 4 and 5. The discussion and conclusions are given in Section 6.

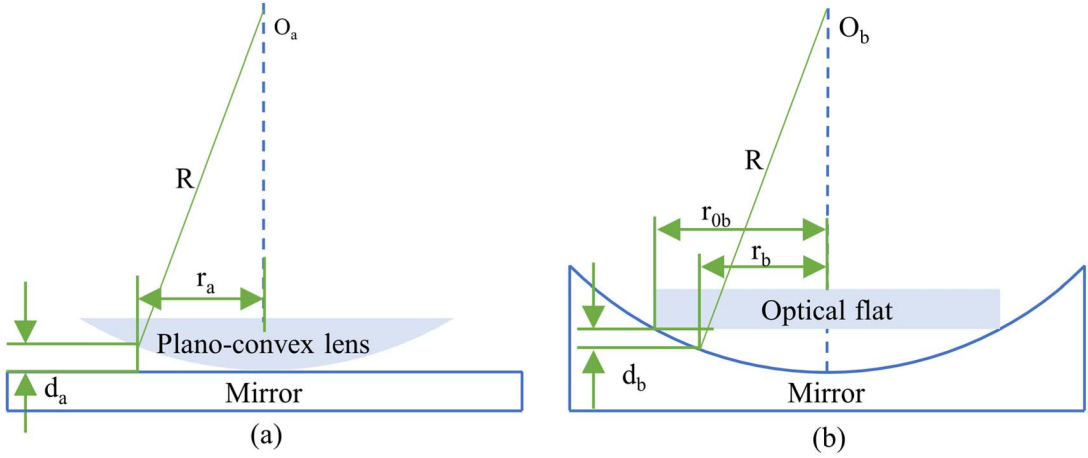

**Figure 2.** (a) Optical structure of Newton rings. (b) The detection principle of the edge sensor based on fringes of equal thickness.

## 2. The Detection Principle, Experimental System, and Mounting Analysis

### 2.1. The Detection Principle of the Edge Sensor Based on Fringes of Equal Thickness

Fringes of equal thickness occur when both coherent beams are obtained by division of the amplitude of the original wave front using a partially reflecting optical surface and then recombined from different paths. Each fringe represents the positions of all points in which two optical surfaces (or wavefronts) have the same path difference. Thus, the detection principles of the Newton ring interferometer and the edge sensor based on fringes of equal thickness are shown in Figure 2.

Consider the optical structure in Figure 2(a), where the spherical surface of one plano-convex lens is placed against the flat surface of another mirror. The initial beam enters the plano-convex lens and partially reflects from the second surface from which the beam is separated into two coherent beams. The coherent beam going through the spherical surface is reflected from the upper surface of the flat mirror. It recombines with the coherent beam reflected from the spherical surface to form fringes of equal thickness. Since the radius of curvature of the spherical surface is known, the spherical surface is the reference surface, and the flat surface is to be measured. Each fringe is a circle, which is called Newton ring. The method of optical testing using interference fringes is widely used in various fields, such as the Fizeau interferometer (de Groot 2014), the Twyman-Green interferometer, and the Mach-Zehnder interferometer (Reichelt & Zappe 2005).

Suppose the shape of the surface to be measured is spherical, and the reference surface is flat. In that case, the fringes of equal thickness will also be a circle after recombining the two beams. The fringe represents the position of all points with equal optical path differences between the two surfaces, whose optical structure is shown in Figure 2(b). $R$ is the radius of curvature of the spherical mirror; $r_{0_b}$ is the radius of the 0-level interference fringe and also the radius of the reference surface; $d_b$ is the thickness of the air gap at a point on the mirror surface, and $r_b$ is the radius of the fringe there. The wavelength of the incident light is $\lambda$. From the geometrical relationship and the equation of optical path difference:

$$d_b = \frac{r_{0b}^2 - r_b^2}{2R} \quad (1)$$

$$\delta = 2nd - \frac{\lambda}{2} = \begin{cases} k\lambda, \\ \frac{\lambda}{2}(2k-1) \end{cases}, k = 0, 1, 2,\ldots \quad (2)$$

Substituting Equation (1) into Equation (2) follows:

$$r_b = \sqrt{r_{0b}^2 - 2d_b R} = \sqrt{r_{0b}^2 - \frac{(\delta + 0.5\lambda)R}{n}}$$

$$= \begin{cases} \sqrt{r_{0b}^2 - \frac{(2k+1)R\lambda}{2n}} \\ \sqrt{r_{0b}^2 - \frac{kR\lambda}{n}} \end{cases}, k = 0, 1, 2,\ldots \quad (3)$$

$n$ in the equation is the refractive index, and $k$ is the fringe level. So the fringe radius can be calculated by the above equations.

Based on the above principle, if two segmented spherical mirrors replace the spherical mirror in Figure 2(b), and the reference flat is set up on the gap of the two segments, the circular interference fringe will be divided into two semicircles. When the two segments are in the co-focus state, the rings of two parts should be concentric; furthermore, when the two segments are in the co-phase state, both sides of the ring are perfectly aligned, the circle centers are coincident, and the radii are equal; the tilt/tip and piston errors of the two segments can be calculated by measuring the fringes' circle centers and radii.

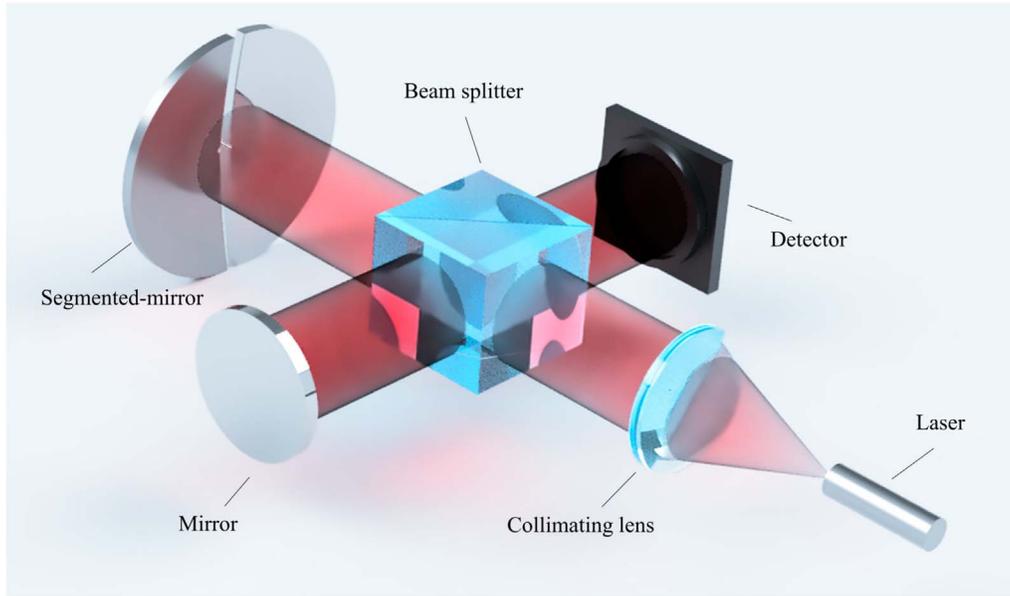

**Figure 3.** Experimental prototype of the edge sensor based on fringes of equal thickness.

Then the position and posture of the segments can be adjusted using actuators to achieve co-focus and co-phase.

In the traditional detection of fringes of equal thickness, the piston error may also be an integer multiple of half wavelength. From Equation (3), it can be seen that the fringe radius is related to the incident light wavelength $\lambda$. Therefore, using two or three lasers (such as the 511, 532, and 632.8 nm) can obtain fringes with different radius sizes. When switching light sources with different wavelengths, the radius of fringes on both sides can be consistent, so the piston error of segments can be measured as 0. In that case, the uncertainty which may exist in the alignment of arbitrary circular fringes can be eliminated by switching the incident light source, and the detection range of piston error can be expanded.

### 2.2. Experimental System Design

An experimental prototype of the edge sensor based on fringes of equal thickness is designed to test the performance and accuracy of the sensor, shown in Figure 3.

The optical system is simple in structure and easy to analyze each surface error in the experiment; the measurement path is separated from the reference path, which can effectively reduce the noise of parasitic reflected light. The experimental system in the laboratory is shown in Figure 4, in which the segmented primary mirror is composed of two spherical hexagonal segments, one of which is a fixed mirror and the other one is a movable mirror. Three high-accuracy actuators are placed under the movable mirror, allowing free adjustment of the mirror position and posture. The edge sensor is placed on the gap between the two mirror segments to detect co-focus errors.

ZEMAX software is used for optical design, mainly to evaluate beam quality, stripe contrast, light intensity distribution, stray light, noise, and system optimization. The design includes light wavelength, optical component size, coating, detector size, and other parameters. The optical system parameters are shown in Table 1.

For this optical system, the theoretical sensitivity of the sensor is $3.''868\,\text{pixel}^{-1}$, and the allowable error of sensitivity is $\pm 0.''232\,\text{pixel}^{-1}$. To achieve the co-focus of all segments, the edge sensor must detect the interference fringe changes when each segment is tilted/tipped (tilt/tip angle with the accuracy of $0.''02$ rms). The design targets of the edge sensors are shown below: the measurement accuracy is $0.''02$ rms, and the measurement range is $\pm 60''$.

### 2.3. Mounting Analysis for Application in LAMOST

Since the sensor system has a simple structure with few components, it is easy to integrate and mount. The preliminary design is shown in Figure 5, the sensor bracket is divided into two parts, the upper bracket is fixed to the outer of the sensor by clamps and connected to the lower bracket through the mirror gap, while the lower bracket is fixed to the mirror chamber structure. In order to minimize the measurement noise and avoid damage to the mirror surface, the sensor body is pressed on the mirror surface with a soft gasket. As the mirror surface is tipped or tilted, the expansion and contraction of the soft gasket can reduce the influence of stray light and dust.

As the edge sensor is mounted in front of the mirror surface, it will block part of the beam. The sensor's size can be minimized by using smaller components. The diameter of the Research in Astronomy and Astrophysics, 23:065022 (14pp), 2023 June                                                                                                                                     Zhang et al.

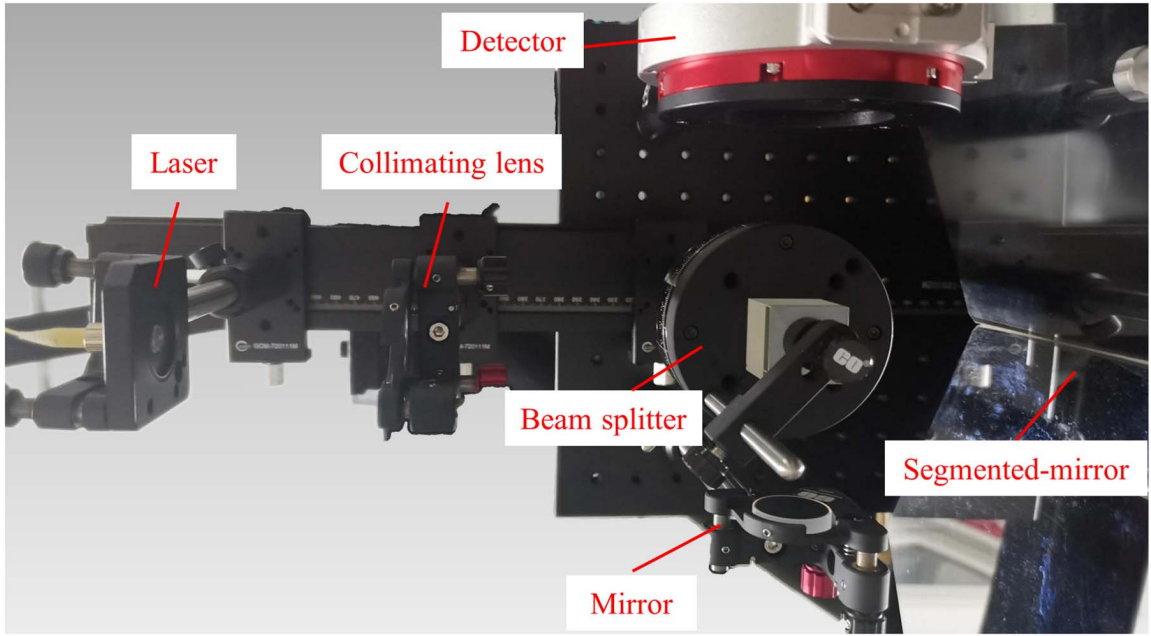

**Figure 4.** Experimental system diagram of equal thickness interference edge sensor.

**Table 1**
Parameters for the Edge Detection System Composed of Two Segmented-mirrors

| Incident Wavelength(nm) | Curvature Radius(mm) | Detector Size (pixel*pixel) | Pixel Size($\mu$m) |
|---|---|---|---|
| 635 | 3000 | 9576*6388 | 3.76 |

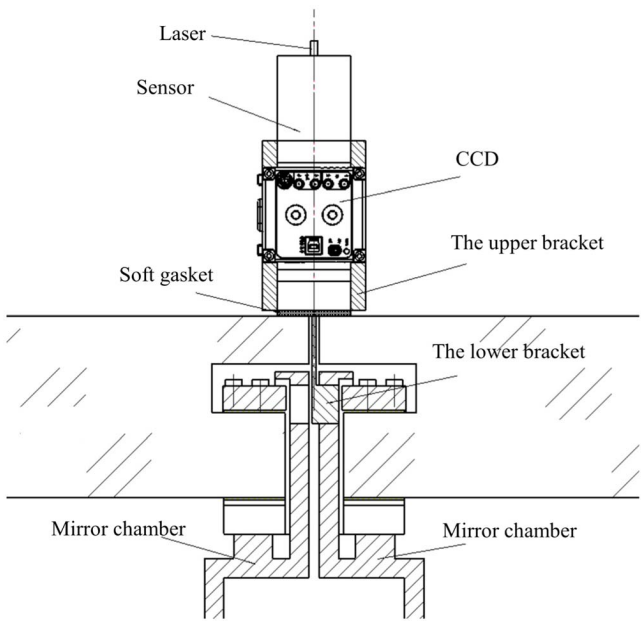

**Figure 5.** The bracket structure of the edge sensor.

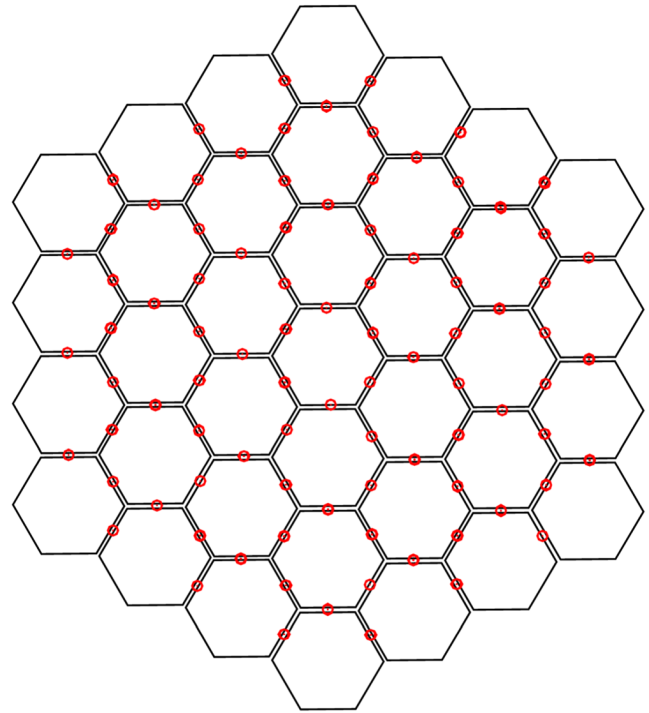

**Figure 6.** Schematic diagram of the edge sensor mounting on the LAMOST.

optical components in the experiment is 25.4 mm, and the detector size is 90 × 97 mm. Therefore, the area blocked by a sensor is 0.00873 m$^2$. Figure 6 shows the mounting of the sensors on the LAMOST, which requires 90 sensors. The total beam blocking area is 0.7857 m$^2$, and the blocking ratio is only 2.7%, which is within the permissible range for the observation with a minimal impact.

## 3. Image Processing Algorithms for Interferograms

The image processing algorithm focuses on detecting the circle center coordinates of the interference fringes, from which the relative tip/tilt error between the neighboring segments can be calculated. The image processing is divided into two steps: image preprocessing and extracting the circle center coordinates of interference fringes using concentric fringe's detection algorithms.

### 3.1. Image Preprocessing

Image preprocessing aims to extract regions of interest (ROI), remove noise, and enhance useful information to facilitate deeper image processing. Therefore, image preprocessing is also divided into two steps: one is ROI extraction, and the other is image enhancement.

The interference fringes of equal thickness in this edge sensor occur by recombination of two coherent beams, one is reflected from a spherical mirror, and the other is reflected by a flat mirror, the beam diameters of the two beams after reflection are different in size. The convergence effect of the spherical mirror makes the reflected beam diameter smaller than the other one. Interference fringes appear only in the range of the smaller beam. Thus, by analyzing the light intensity changes of the image, the position of the smaller beam can be located so that ROI extraction is realized.

Image enhancement can be divided into two methods: the frequency domain method and the spatial domain method (Gonzal & Woods 2002). In this process, combining two methods can obtain a better result. A high-pass filter in Fourier space is used to remove low-frequency noise. Then, various spatial domain methods (nonlinear filtering, contrast limited histogram equalization (Wang & Ward 2007), adaptive binarization (Sauvola & Pietikäinen 2000), etc.) are used to deal with other noise. Through this process, feature regions are highlighted, other regions are suppressed, and the clarity of the valid information in the image is enhanced to meet the subsequent processing needs.

The ideal and noise-containing interferograms are shown in Figures 7(a) and (b). Figures 7(c) and (d) show the original interferogram obtained in the experiment and the interferogram after preprocessing.

### 3.2. Concentric Fringes Detection Algorithms for Interferograms

Since the interference fringes are concentric fringes composed of many bright and dark circles, circle detection is vital to the concentric fringe's detection algorithm. Circle detection methods widely used can be divided into two categories: pixel-based methods and contour-based methods (Gong et al. 2018; Martorell et al. 2021). The pixel-based method mainly detects by Hough transform (HT) or its variants to be classified and extracts the image's pixels, which is robust to all kinds of image detection (Hough 1962). However, the detection speed of this method is slow because of the large amount of storage space required. Contour-based methods, also known as edge-tracking algorithms, detect geometric structures by tracking the connectivity between pixels. Using contours rather than individual pixels can significantly improve the computational speed. The structure of the concentric fringe's detection algorithm based on interference fringe contours is shown in Figure 8.

After image preprocessing, a contour data set consisting of interference fringe contours can be obtained using a contour tracking algorithm. The circle center coordinates of each contour are fitted using the least square method to form the set of circle center points. The standard deviation of the circle center points set $O_\sigma$ is calculated. The clustering degree of the circle center set is determined by comparing the magnitude of the empirical threshold $P$ with $O_\sigma$: if $P < O_\sigma$, there are outliers with significant deviation, then an empirical threshold $Q$ is used to determine the outliers. After the outliers are deleted, a concentric circle candidate set can be obtained, and its clustering degree is determined again until it meets the requirement. In the end, the circle center set with a reasonable clustering degree is regarded as the concentric circle set, and its median is the detection result.

Comparing the size and clustering degree of the circle center set and the execution time, the above two types of algorithms can be evaluated, as shown in Figure 9. It can be found that the pixel-based algorithm obtains a larger amount of data on the circle centers and gets more data points than the contour-based algorithm. Hence, there are poor clustering and more outliers, which need to be repeated several times for filtering. Compared with the pixel-based algorithm, the contour-based algorithm has higher detection accuracy and shorter execution time.

## 4. Simulation Tests

Co-focus simulations are carried out by processing the interferograms generated by the simulation software to evaluate the performance of the edge sensor, including the accuracy, sensitivity, linearity, and measurement range.

ZEMAX software is used to generate the ideal interferograms for the system. Noise-containing interferograms can be obtained by adding random noise to the ideal interferograms.

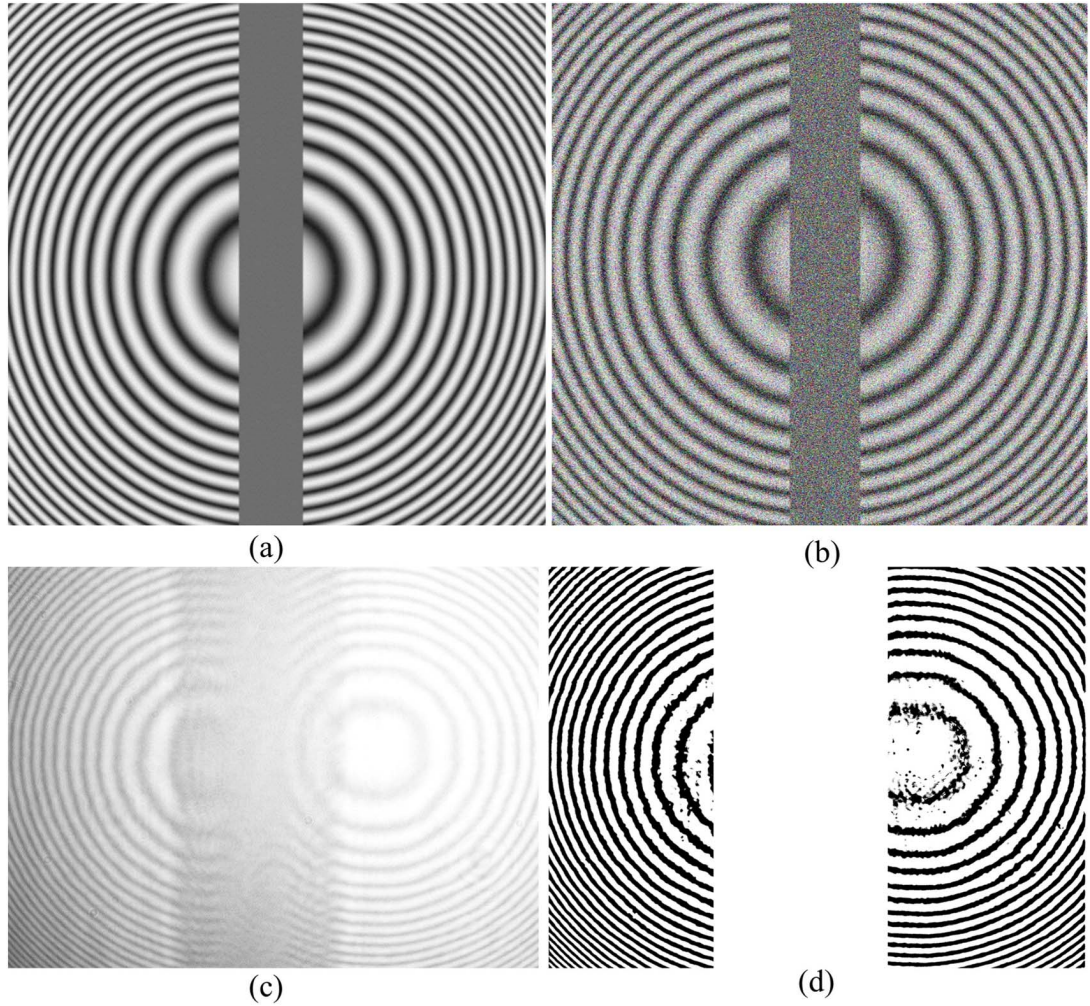

**Figure 7.** (a) The ideal interferogram in simulation. (b) The simulation interferogram with noise. (c) The experiment interferogram. (d) The experiment interferogram after preprocessing.

The robustness of the image processing algorithm can be evaluated under different noise conditions by examining the ideal and noise-containing interferograms separately.

### 4.1. Simulation Tests of Ideal Interferograms

In the first step, the accuracy and sensitivity of the edge sensor can be tested. The segmented mirror system has two segments, one is fixed, and the other is movable. The tilt angle of the movable segment is changed gradually from $0''$ to $0''\!.4$ in steps of $0''\!.01$, and 40 interferograms can be obtained. In the same way, gradually changing the movable segment tip angle can also obtain 40 interferograms. The accuracy and sensitivity of the edge sensor can be achieved by analyzing the changing regularity of the concentric coordinates.

The relationship between the tilt angle and the $y$-axis coordinates of fringe centers is shown in Figure 10. When the mirror's tilt angle changes, the fringe center shifts with it, and the $y$-axis coordinate of the fringe center is linearly related to the tilt angle. The maximum error in the $y$-direction is 0.00279 pixel ($0''\!.00072$). A linear fit to the $y$-axis coordinate data of the circle centers is performed, and the fitted equation slope is the sensitivity of the edge sensor in the tilt direction. It can be seen that the sensitivity in the tilt direction is $3''\!.852\,09 \pm 0''\!.00166\ \text{pixel}^{-1}$.

Similar simulation results in the tip direction are shown in Figure 11. It can be seen that when the movable mirror's tip angle changes, the fringe center's $x$-axis coordinate changes linearly. The maximum error in the $x$-direction is 0.0043 pixel) ($0''\!.0011$), and the sensitivity in the tip direction is $3''\!.893\,06 \pm 0''\!.00527\ \text{pixel}^{-1}$.

In the next step, the linear measurement range of the sensor is calculated. The tilt/tip angle of the movable segment is separately changed gradually from 0″ to 100″ in steps of 10″, and two sets of interferograms can also be obtained. Increasing the tilt/tip angle can measure its linear operating interval; analyzing the deviation between the measured and theoretical values can evaluate the linearity of the sensor within the measurement range. Therefore, when the mirror is gradually tilted/tipped in steps of 10″, the changes of the fringe center coordinates in the tilt/tip direction are shown in Figure 12(a)/Figure 12(b), and the errors are shown in Figure 12(c).

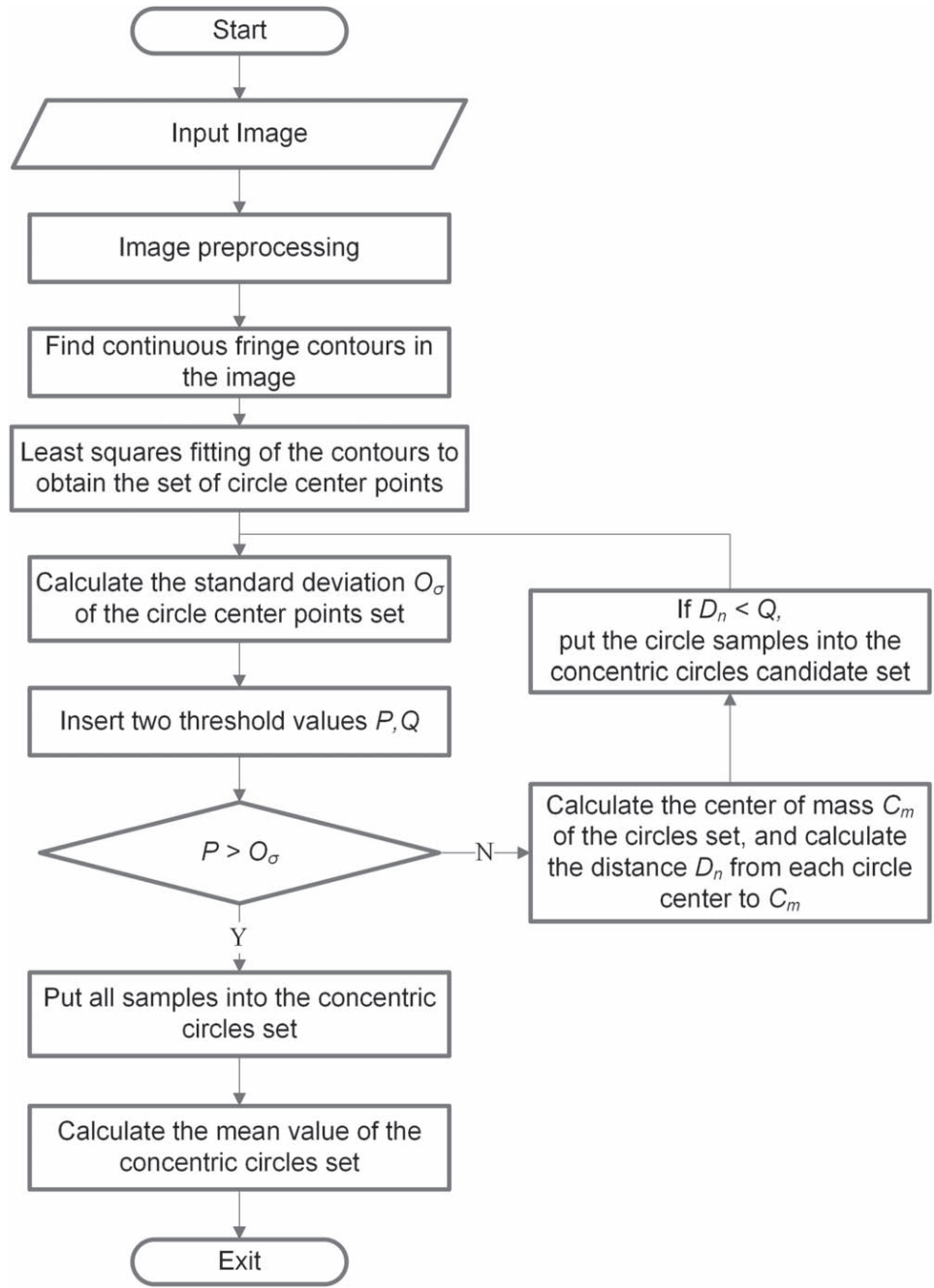

**Figure 8.** Structure of the concentric fringe's detection algorithm based on interference fringe contours.

Figures 12(a) and (b) show that the slope values of both lines obtained by linear fitting are compatible with the sensitivity requirements. The fitted straight lines overlapped very well with the theoretical ones. It should be noted that the absolute error in Figure 12(c) is caused by the image detection algorithm. The absolute errors in the tilt direction are smaller than those in the tip direction significantly, which is caused by the interferograms. Since the sensor is located in the gap of two mirror segments, the interferogram is split into two semicircles along the y-axis. When the x-axis coordinate is detected, the semicircle is located on the side of the circle center, so the result slightly deviates. In contrast, when the y-axis coordinate is detected, the semicircle is symmetrical, so the detection result is closer to the theoretical value.

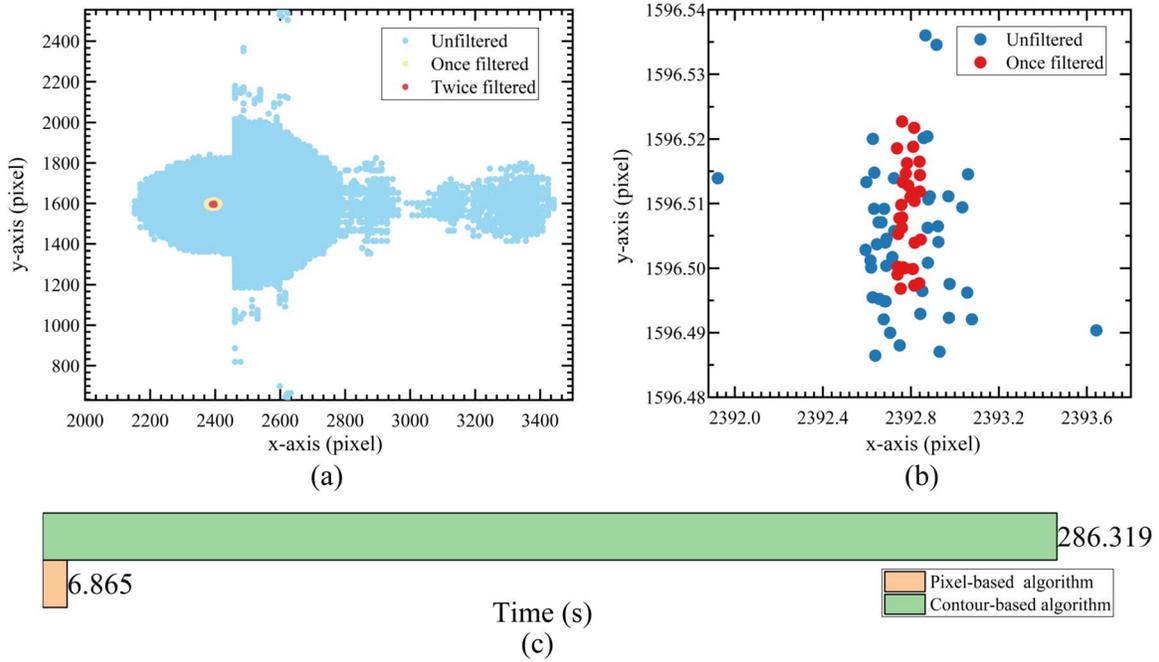

Figure 9. (a) Circle center set results from the pixel-based detection algorithm. (b) Circle center set results from the contour-based detection algorithm. (c) Execution time of the two algorithms.

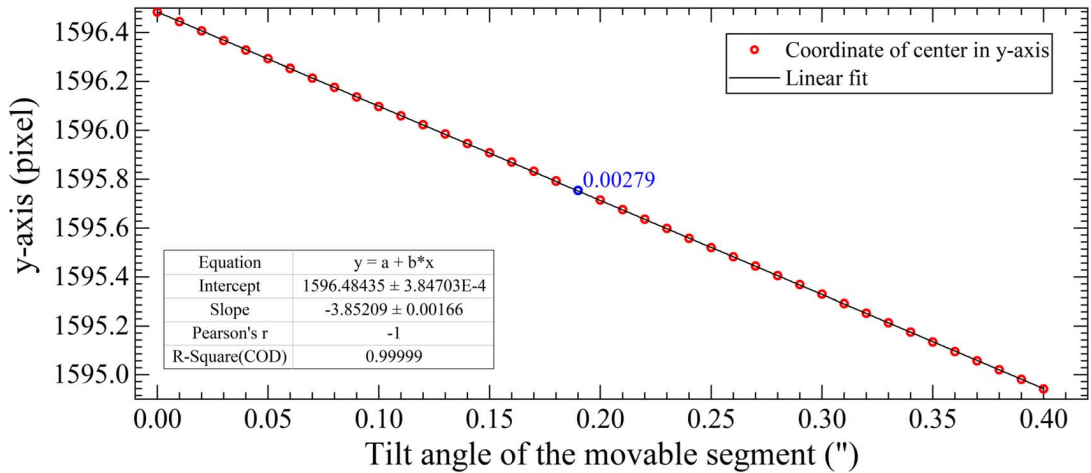

Figure 10. The relationship between the tilt angle and the y-axis coordinates of the fringe center.

However, the error values of the detection results on both sides are within the allowed range. Figure 12(c) shows that when the tilt/tip ⩽100″, the maximum error in the tilt direction is 0.00886 pixel, and the maximum error in the tip direction is 0.07277 pixel. Both are well within the accuracy of 0.″02 rms (Root Mean Square, 0.077 pixel) or 0.″12 PV (Peak to Valley, 0.464 pixel). Linearity error is the sensor output curve deviation from a specified theoretical straight line over the

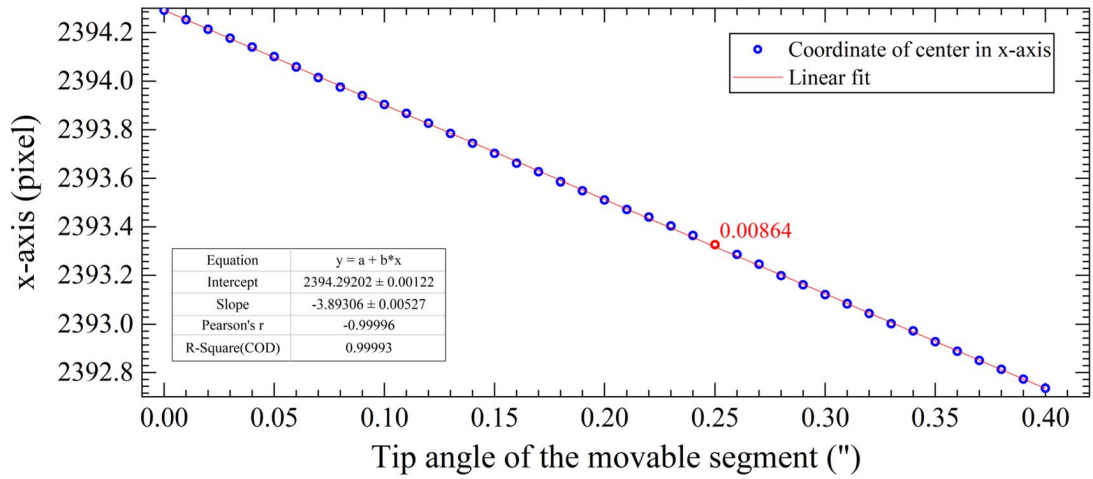

**Figure 11.** The relationship between the tip angle and the *x*-axis coordinates of the fringe center.

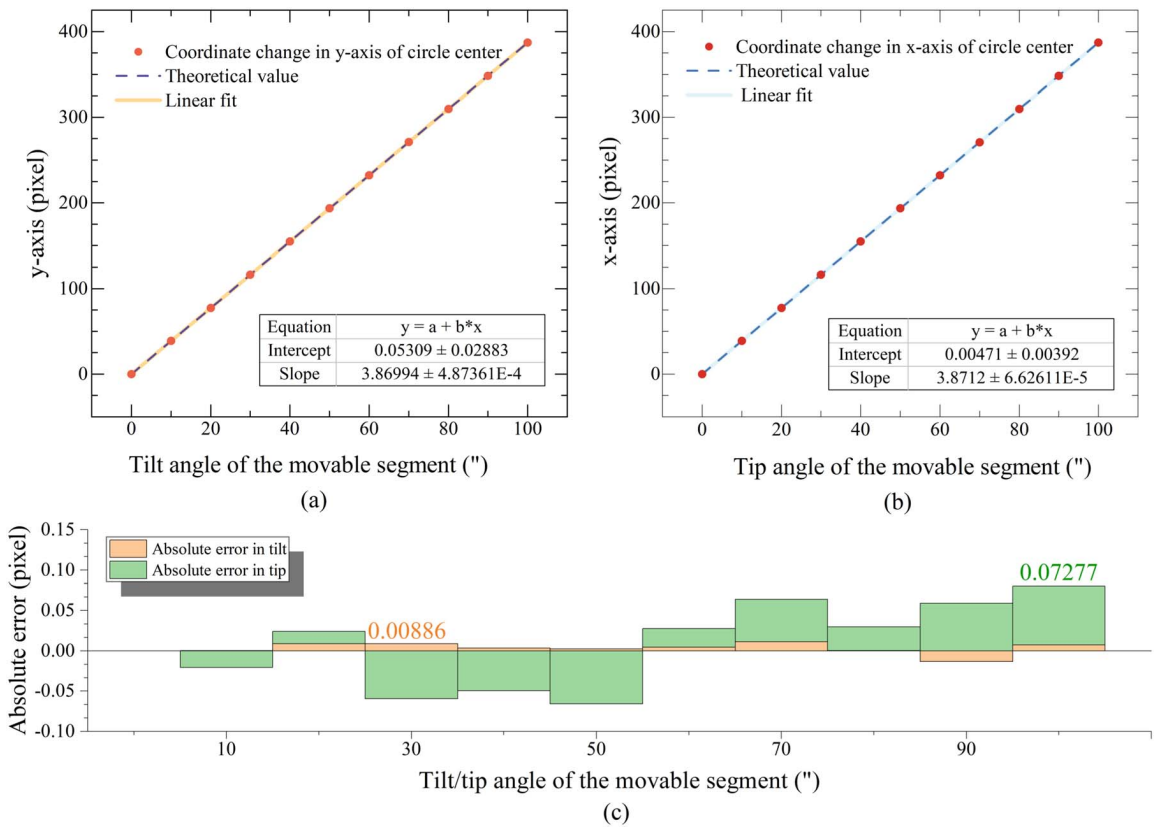

**Figure 12.** (a) *Y*-axis coordinate changes of fringe centers when the tilt angle changes from 0″ to 100″. (b) *X*-axis coordinate changes of fringe centers when the tip angle changes from 0″ to 100″. (c) Absolute tip/tilt errors.

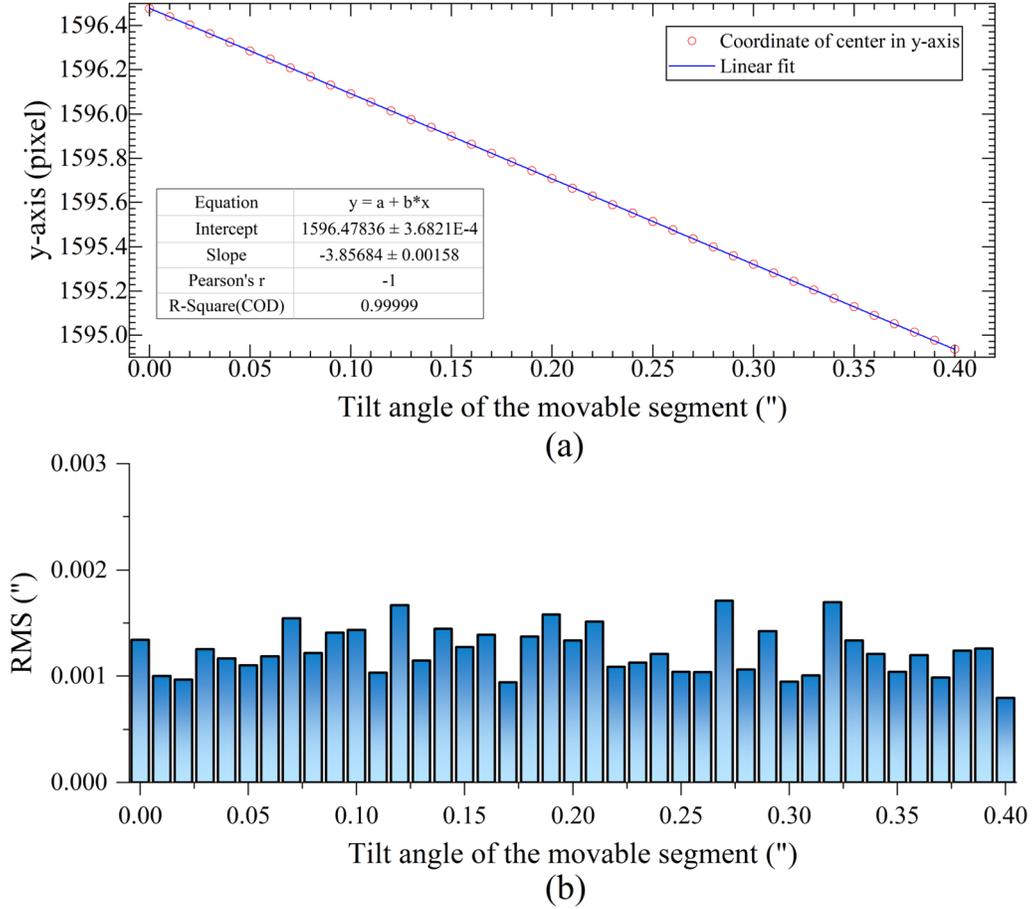

**Figure 13.** (a) The relationship between the tilt angle and the *y*-axis coordinates of the fringe center to the noise-containing interferograms. (b) The rms of the tilt detection.

desired range. The equation of linearity error is given below:

$$\delta = \pm \frac{\Delta Y_{max}}{Y_F} \times 100\% \quad (4)$$

$\Delta Y_{max}$ is the maximum deviation (pixel) of the measured data to the theoretical straight line, which is the maximum absolute error; $Y_F$ is the measured value (pixel) of the sensor at full range. From this, the linearity of the sensor can be calculated separately in the interval of tilt/tip $\leqslant 100''$:

$$\begin{cases} \delta_{tilt} = \frac{+0.00886}{386.88095} \times 100\% = +0.00229\% \\ \delta_{tip} = \frac{+0.07277}{386.88095} \times 100\% = +0.01881\% \end{cases} \quad (5)$$

In actual application, the range of the edge sensor is not only related to its detection capability but also limited by the actuator movement range. Moreover, the tilt/tip angle error is usually on the order of tens arcsec scale for the segments, so this measurement range can fully meet the co-focus requirement.

### 4.2. Simulation Tests of Noise-containing Interferograms

In this section, the two sets of interferograms from Section 4.1 are processed by several random noise algorithms, such as salt and pepper noise, Gaussian noise, and others. After adding noise, two sets of noise-containing interferograms are obtained. The SNR of the processed noise images is about 20 dB. Since the noise-adding algorithm's random effect influences the measurement accuracy, we use a signal-averaging method to remove it: an interferogram is processed 25 times by the random noise-adding algorithm to obtain a set of noise-containing interferograms, and the final result is obtained by averaging the results of all detections. The relationship between the fringe center coordinates of the noise-containing interferograms and the tilt/tip angle changes can be obtained, as shown in Figures 13(a) and 14(a). The accuracy is represented by the rms, which is shown in Figures 13(b) and 14(b).

It is shown that in noise-containing interferograms (SNR is 20 dB): the detection sensitivity in the tilt direction is $3''85684 \pm 0''00158$ pixel$^{-1}$, and the rms of all detected results are less than 0.002;" the sensitivity in the tip direction detection is $3''86342 \pm 0''0051$ pixel$^{-1}$, and the rms of all detected results are less than $0''006$. Therefore, when the SNR of interferograms is larger than 20 dB, the edge sensor can still meet the demand for segmented mirror co-focus detection, and the detection accuracy is much better than $0''02$ rms.

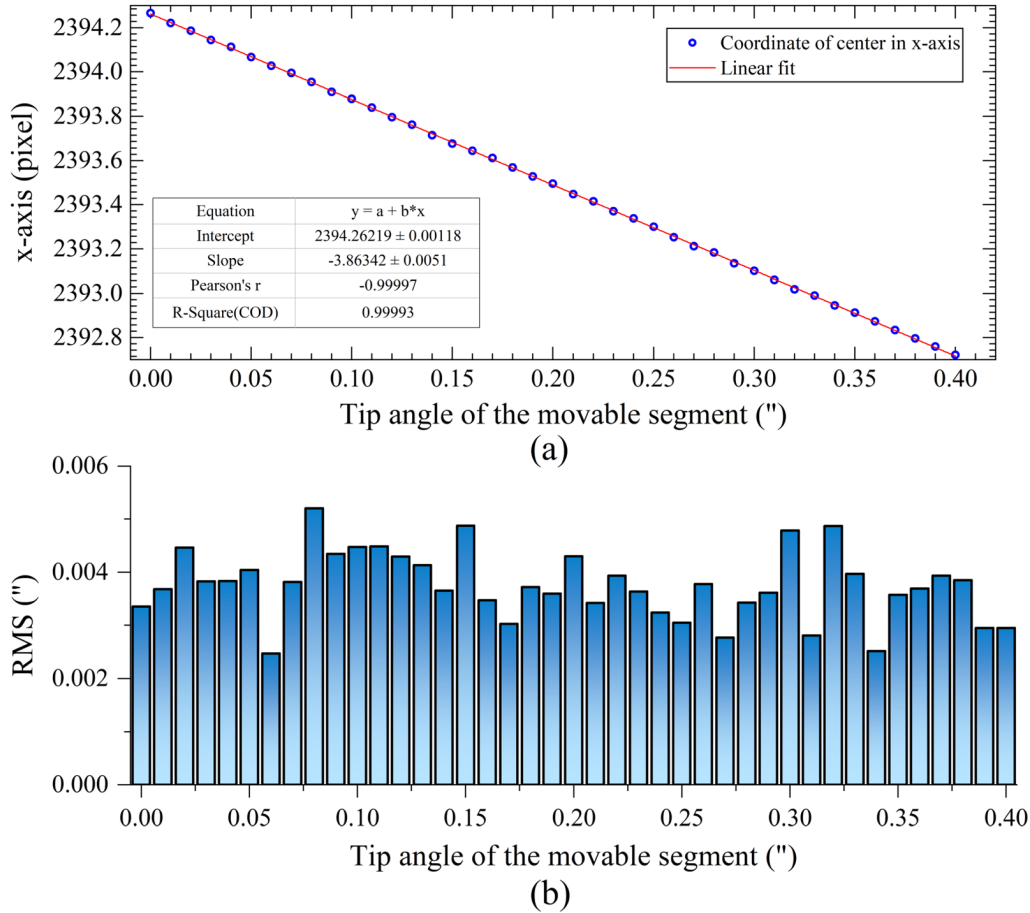

**Figure 14.** (a) The relationship between the tip angle and the *x*-axis coordinates of the fringe center to the noise-containing interferograms. (b) The rms of the tip detection.

## 5. Experimental Test

The repeatability, sensitivity, linearity, and accuracy of this edge sensor were tested in the experiment. The systems in the laboratory consisted of a fixed mirror and a movable mirror, where three high-precision actuators were installed under the movable mirror. The actuators A1, A2 and A3 are shown in Figure 15. According to the geometric relationship, when the actuator A2 moves, the movable mirror rotates around the line A1A3, and the tip angle of the mirror changes; if the actuator A2 moves in 1 $\mu$m, the relative tip angle of the movable mirror is $1''62$.

In practical applications, the repeatability of a sensor is an important parameter. The measurement results' rms can be used to represent the repeatability quantitatively. In addition, higher measurement accuracy can be achieved if the average of all measurements is used as the final measurement result. In this regard, we designed the following experiments to test the edge sensor's repeatability and find the optimal number of repeated measurements to meet the design targets.

First, 100 interferograms are taken and detected with 20 nm ($0''0324$) as the actuator A2 movement step. Second, ten interferograms were taken as a group, and the average value of the ten images was taken as the measurement result of the group. Finally, the ten groups' rms is calculated. According to those methods, the number of repeated measurements for a group of images is set to 10, 15, 20, 25, 30, and 35. Figure 16 shows the distribution of rms and PV values of the measurement results for the six sets.

It can be seen that with the increase in the number of measurements, the trend of rms results first decreases and then increases. It means that an appropriate increase in the number of repeated measures does help to improve the measurement accuracy. Still, as the measurement time increases, it is possible to introduce other errors that lead to a decrease in measurement accuracy. Figure 16 shows that the minimum rms is at the number of repeated measurements in 25.

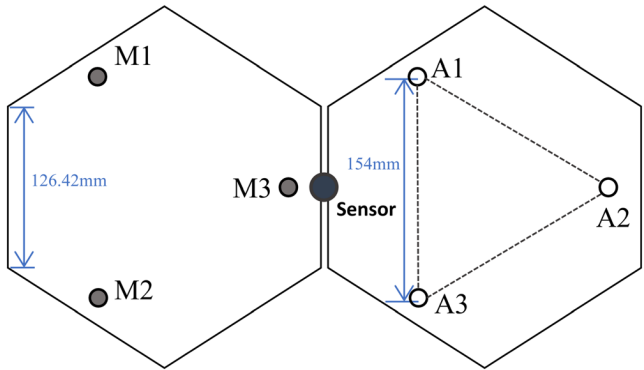

**Figure 15.** Geometrical arrangement of actuators A1-A3 and the edge sensor.

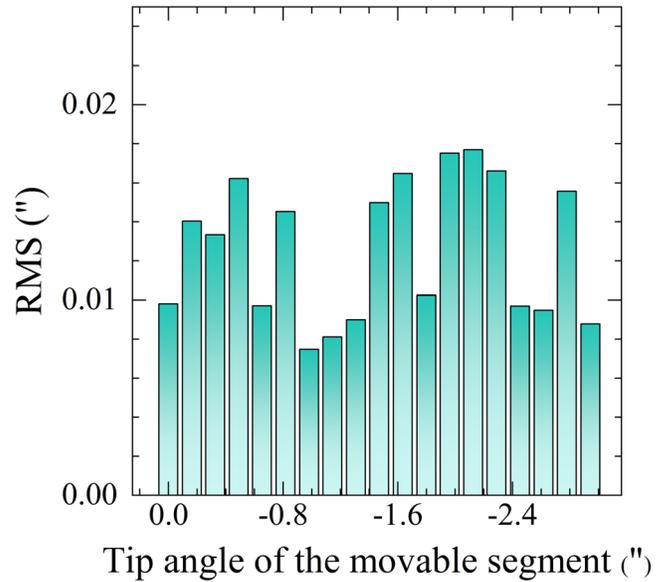

**Figure 18.** The rms for each measurement.

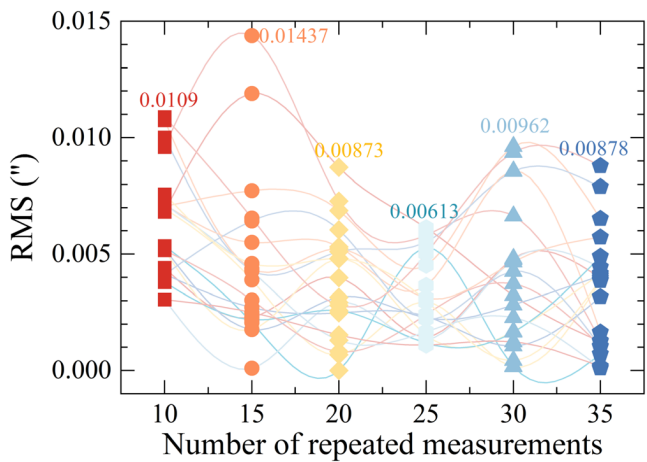

**Figure 16.** The rms in repeated measurements.

In the next experiment, the step of tip angle changes is set to a fixed value, which can detect the sensitivity in the fine-tuning measurement range. The movement of actuator A2 is 100 nm one time, corresponding to $0\rlap{.}''162$ in tip angle, and 25 interferograms are taken for each move. The detection results are shown in Figure 17, and the rms of each measurement is shown in Figure 18.

The experimental results showed that the sensitivity is $3\rlap{.}''8946\pm0\rlap{.}''0277\,\text{pixel}^{-1}$, and the absolute error is 0.17038 pixel. Based on the maximum absolute error, the linearity is about 1.48%. Both rms is less than $0\rlap{.}''02$. Therefore, this edge sensor can detect the tilt/tip error of the segmented mirror, and all relevant performance parameters meet the design target.

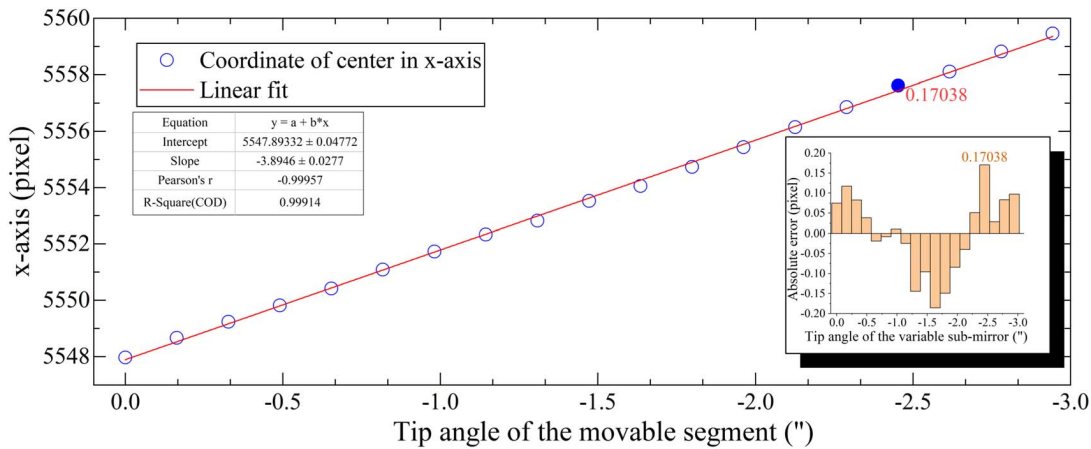

**Figure 17.** *X*-axis coordinates of the fringe center on the experimental interferograms when the actuator A2 is moved downward with a step in 100 nm.

## 6. Conclusion

Co-focus and co-phase detection is one of the key technologies for large aperture segmented mirror telescopes. Based on the fringes of equal thickness and the two-segmented mirror active optics experiment system, we successfully developed a new type of edge sensor prototype, together with two kinds of concentric fringes detection algorithms. Simulations and experiments have shown that the edge sensor can achieve adequate accuracy better than $0.''02$ rms and the linearity is less than 1.48%. Such results have met the co-focus detection requirements of most large or extremely large segmented mirror astronomical telescopes.

According to different segmented mirror requirements, we should only adjust and optimize the sensor parameters, such as the surface shape of the segmented mirror, the wavelength of light, the diameter of the beam, the diaphragm aperture, the pixel size and resolution of the detector, to achieve the optimal system performance and lower manufacture cost. Because of the above advantages and avoiding environmental factors influences and on-sky calibration demand from most electrical edge sensors, it is feasible and low-cost to build this optical edge sensor based on fringes of equal thickness for most segmented mirror telescopes.

In the future, this sensor will be used on LAMOST for co-focus testing. Further engineering research will be carried out for practical applications in LAMOST. In addition, to achieve the co-phase of the segmented mirrors, the piston error of the segments needs to be detected. It is feasible to detect the radius when the center position of the fringes of equal thickness is known. Based on the concentric fringes detection algorithm mentioned in this paper, we will develop another algorithm to detect the radius of those concentric fringes. Piston error detection using this edge sensor is planned to be tested and studied so that this kind of edge sensor could be adapted to the co-phasing process of the segmented telescope. Therefore, the edge sensor could be used for large telescopes in the future, such as the 12 m Large aperture Optical/infrared Telescope (LOT) in China.

## Acknowledgments

Guo Shou Jing Telescope (the Large Sky Area Multi-Object Fiber Spectroscopic Telescope) is a National Major Scientific Project built by the Chinese Academy of Sciences. Funding for the project has been provided by the National Development and Reform Commission. LAMOST is operated and managed by the National Astronomical Observatories, Chinese Academy of Sciences. This work is supported by the National Key R&D Program of China (Grant Nos. 2022YFA1603002, 2022YFA1603001, 2021YFC2801402 and SQ2021YFC2800011) and the National Natural Science Foundation of China (Grant Nos. U2031207, U1931207, 12073053, and 1331204).

## ORCID iDs

Xi Zhang  https://orcid.org/0000-0001-9128-9636
Heng Zuo  https://orcid.org/0000-0001-9764-2082
Yong Zhang  https://orcid.org/0000-0003-2179-3698